# Diffusion Model-based Probabilistic Downscaling for 180-year East Asian Climate Reconstruction

**Authors:** Fenghua Ling[1,2†], Zeyu Lu[3,4†], Jing-Jia Luo[1*], Lei Bai[3*], Swadhin K. Behera[2], Dachao Jin[1], Baoxiang Pan[5], Huidong Jiang[6,7] and Toshio Yamagata[1,2]

**Affiliations:**
[1]Institute for Climate and Application Research (ICAR)/CIC-FEMD/KLME/ILCEC, Nanjing University of Information Science and Technology, Nanjing, China
[2]Application Laboratory, Japan Agency for Marine-Earth Science and Technology, Yokohama, Japan
[3]Shanghai AI Laboratory, Shanghai, China
[4]Shanghai Jiao Tong University
[5]Institute of Atmospheric Physics, Chinese Academy of Sciences, Beijing 100029, China
[6] Department of Computer Science, Tokyo Institute of Technology, Yokohama, Japan
[7]RIKEN Center for Advanced Intelligence Project, Tokyo, Japan

†Equal Contributions,
*Corresponding to jjluo@nuist.edu.cn, bailei@pjlab.org.cn

**ABSTRACT**

As our planet is entering into the "global boiling" era, understanding regional climate change becomes imperative. Effective downscaling methods that provide localized insights are crucial for this target. Traditional approaches, including computationally-demanding regional dynamical models or statistical downscaling frameworks, are often susceptible to the influence of downscaling uncertainty. Here, we address these limitations by introducing a diffusion probabilistic downscaling model (DPDM) into the meteorological field. This model can efficiently transform data from 1° to 0.1° resolution. Compared with deterministic downscaling schemes, it not only has more accurate local details, but also can generate a large number of ensemble members based on probability distribution sampling to evaluate the uncertainty of downscaling. Additionally, we apply the model to generate a 180-year dataset of monthly surface variables in East Asia, offering a more detailed perspective for understanding local scale climate change over the past centuries.

**INTRODUCTION**

As the average global temperature continues to rise and extreme weather events become more frequent, the impact of climate change on our lives is increasingly apparent[1,2]. However, the rate of warming varies across regions, as does the frequency of such extreme events[3–6]. Therefore, reliable and accurate regional climate information is crucial to deal with local climate change and its impacts. While many international efforts have been put to develop high-resolution global climate models (GCMs) in recent decades[7,8], it is important to note that when assessing century-scale climate changes, most GCMs still operate with grid spacings of 100 km or more due to limitations in computational resources[9–11]. Such coarse spatial resolutions have proved to be inadequate for evaluating climate changes at the scale relevant to local communities[12]. As a result, this challenge has given rise to the adoption of downscaling techniques, which are now widely employed to bridge the gap between global climate projections and the specific climate information needs of local communities[9,13–16].

These downscaling techniques, including both dynamical and statistical methods, have been developed to generate high-resolution climate data[13–20]. Among these techniques, Regional Climate Models (RCMs) stand out as dynamical models that utilize topography and circulation conditions from GCMs to generate the regional climate information[13–15]. However, it is important to note that RCM downscaling results are reliant on the large-scale circulation conditions provided by the GCMs, which leads to the propagation of systematic biases from the GCMs and an increase in uncertainties in the downscaled result[21]. Furthermore, dynamical models incur significant computational expenses and necessitate substantial data storage and processing costs [13–15]. In contrast, statistical downscaling methods can provide high-resolution outputs akin to dynamical downscaling but with significantly reduced resource and computational demands. Statistical downscaling methods leverage statistical relationships between low-resolution and high-resolution climate data. These statistical downscaling methods rely on mathematical techniques, including deep learning and traditional statistical approaches, to establish statistical relationships between low-resolution and high-resolution climate data, enabling the derivation of detailed local-scale information[17–20,22]. Currently, both dynamical and statistical downscaling techniques find widespread use in studies related to climate change, climate variability, hydro-climate extremes, and impact assessments at regional scales, particularly within sectors such as agriculture, energy, and water resources[23–26].

Unfortunately, both dynamical and statistical downscaling methods currently prioritize deterministic modelling, which often leads to the oversight of the inherent uncertainties within the data and the ill-posed problem of the downscaling process[27,28]. Those uncertainties have become a growing concern, as precise estimations of climate change and robust assessment methods are crucial for a comprehensive understanding of climate change. Many are now exploring innovative technologies and approaches to tackle the challenges posed by climate change[29–32]. Here, we introduce the Diffusion Probabilistic Downscaling Model (DPDM), a data-driven approach designed to simulate the probability distribution function of high-resolution climate based on the corresponding low-resolution data. Furthermore, the DPDM incorporates a conditional probability method that accounts for the influence of external factors, such as topography information during the downscaling process. Compared with deterministic downscaling, DPDM derives probability distribution functions and generates a large number of ensemble members[33], which not only obtains the accurate local details but also allows for more robust estimates of uncertainties. Additionally, we apply DPDM to the NOAA-20th century reanalysis[34] to offer a detailed information on surface climate over East Asia from 1836 to 2015, which significantly enhance our understanding of local-scale climate changes since the late 19th century.

# RESULTS

## Climate Downscaling via the DPDM

For monthly climate downscaling, one of the main limitations of deep learning methods is the lack of long-term high-resolution observational data. To mitigate this, we use the easily available ERA5-Land dataset[35], which offers the surface meteorological variable data on a high resolution of 0.1°. Furthermore, for a more robust estimation of uncertainty in downscaling, we use the lower-resolution 1° from ERA5 dataset rather than relying on interpolation[36]. The ERA5 and ERA5-Land are produced by different versions of the Integrated Forecast System model, with ERA5-Land uses an improved land surface model that can better represent processes such as surface-atmosphere interactions, soil moisture dynamics, and vegetation processes compared to the one used in ERA5. In addition, in order to ensure sufficient training data, we assume that the mapping relationship between large-scale areas and small-scale details in downscaling is consistent at different time scales. Consequently, we utilize 6-hourly data spanning the period from 1961 to 2015 for model training (details in Table S1 and Methods).

Recognizing that many of the intricate details in high-resolution data are affected by topography information, we challenge to focus on the East Asia region (65°E-135°E, 15°N-55°N), which boasts some of the most complex and diverse terrain conditions. To comprehensively account for the impact of varying terrains, we introduce terrain and land-sea boundary data into the input dataset of DPDM. Furthermore, we employ a training strategy that divides an input image into a series of patches and randomly selects a patch as the training input. This strategy enhances the diversity of training data and reduce training complexity. Besides this, DPDM train with this strategy demonstrates robust generalization capabilities and can easily be adapted to regions beyond East Asia or to smaller areas through simple fine-tuning. In terms of our model's architecture, we are inspired by the well-established model (SR3)[37]. Additionally, to improve the efficiency of data generation process, we apply the sample process from Denoising Diffusion Implicit Model (DDIM[38]), leveraging its flexibility to skip some steps in the denoising process and trade off quality for generation speed, as fewer steps inevitably lead to a loss of quality but 250 steps were deemed sufficient for our purposes (ref. 39, Fig. S1 and Table S2).

Table 1 unequivocally highlights the superiority of DPDM in downscaling results compared to other statistical methods. This includes deterministic models such as Enhanced Deep Residual Networks for Super-Resolution with Generative Adversarial framework (EDSR-GAN), known for their superiority over traditional statistical techniques[40-43], as well as the widely-used linear interpolation methods (Lerp) in meteorology[44,45]. The consistent results across a variety of meteorological data reveal that the DPDM outperforms other models in terms of key metrics, including the Anomaly Correlation Coefficient (ACC), Peak Signal-to-Noise Ratio (PSNR), Structure Similarity Index Measure (SSIM), and Root Mean Square Error (RMSE). It is worth mentioning that this superiority is particularly evident in the case of precipitation and temperature downscaling, which are very important for social life in a warming climate. Furthermore, a distinct advantage of DPDM is the capacity to generate multiple ensemble members by sampling from the probability distribution. We conduct a comparative analysis of these metrics for varying membership sizes, including ensemble mean with 33 and 100 members. The results clearly demonstrate that increasing the membership size enhances downscaling capabilities proportionally. It is worth noting that DPDM with multi-member ensembles will increase the inference time compared to other deterministic models (Fig. S2). Nevertheless, when compared to dynamical regional climate models (RCMs), the computational efficiency of DPDM remains acceptable.

To assess their performance in reproducing the spatial distribution, we compare the patterns of climatological mean precipitation and its variance in the test dataset based on the different downscaling methods (Fig. 1). It is evident that the Lerp yields overly smooth results, failing to accurately reproduce local details (Figs. 1a4-d4). In contrast, both the deterministic and DPDM impressively capture the spatial details of the precipitation distributions and its temporal variability, as evidenced by improved spatial pattern correlation and reduced spatially averaged Normalized Root Mean Square Error (NRMSE, see Figs. 1 and S4).

However, we note that the deterministic method exhibits lower correlation in mid-high latitudes and plateau areas. This divergence can be attributed to the complex and variable land surface characteristics in such terrains, resulting in variations even under the same large-scale background conditions. In addition, methods based on the deterministic approach (EDSR-GAN) sometimes introduce false features[46,47](Figs. 1c2 and 1d2). Interestingly, DPDM, especially that with multiple members, is able to effectively address these issues. This conclusion is parallel with the findings in distribution histograms, where the Lerp struggles to accurately represent data variation ranges, and where EDSR-GAN tends to exhibit a potential for overestimation (see Fig. 1e and Fig. S2). For the downscaling of other meteorological variables, DPDM also exhibits similar advantages over the deterministic model (Fig. S5 and S6).

To evaluate whether the downscaled results effectively capture high-resolution details, we employ an objective assessment based on the R squared ($R^2$) and blurriness. Blurriness serves as a crucial metric for objectively assessing whether a model exhibits excessive smoothing or sharpness, avoiding subjective visual assessments[20]. We utilize the absolute difference between the high-resolution data and the Lerp results to evaluate whether the high-resolution data contains sufficient detail information (shown in Fig. 2 c1-c5) because the Lerp results are smooth enough (Fig. S7). If the data predominantly falls on the left side of the diagonal, it suggests that the downscaled data's information content is less than that of true high-resolution data, indicating a bias toward smoothing. Conversely, if the data leans toward the right side of the diagonal, it signifies excessive sharpness, introducing more information that deviates from reality. Notably, our DPDM with 100 members exhibits a higher degree of model fitting when compared to deterministic models. The DPDM effectively preserves true local details in downscaling results, although some smoothing is visible in precipitation. The evaluation highlights DPDM's capability in reconstructing high-resolution details, even with multiple ensemble members, achieving similar or superior performance compared to other methods (Fig. S8), while also mitigating issues such as excessive smoothing or sharpness observed in other models.

**Uncertainties of DPDM**

The DDPM can generate multi-member downscaling results through probability distribution sampling. Therefore, we evaluate in detail how the number of ensemble members affects the downscaling performance and why they can be used to evaluate uncertainty (Fig. 3).

As the number of members is increased, the ability of DPDM to capture true details improves in all aspects[48]. An interesting feature is that the improvement in downscaling capability from 30 to 100 members is relatively small. This is reminiscent of the conclusion reached by dynamical models using large members for ensemble predictions. Part of the reason for this phenomenon is that approximately 30 members are sufficient to represent most of the high-resolution details and to cover the uncertainty space in the downscaling. The ensemble scheme of the dynamical model is similar to the ensemble scheme of DPDM in that both the methods involve sampling multiple results, although the sampling methods are different. Dynamical models utilize randomly perturbed parameterization or randomly perturbed physical processes to generate multiple results, whereas DPDM samples multi-members from the probability distribution.

We find significant varieties among different members, as shown in the spatial pattern of precipitation in July 2020 in Fig. S9. Fig. 3b highlights the considerable uncertainty in the model's downscaling output. But it is worth noting that multi-member average of the model output will better improve the downscaling capability. Furthermore, we conduct a specific assessment on extreme rainfall events. In July 2020, severe floods occurred in the Yangtze River Basin in East Asia, and drought occurred in southern China (Fig. 3c and 3d). In the two extreme cases, we find that the results obtained through multi-member mean ensemble scheme are always closer to the true values than the Lerp and single-member results, no matter whether dealing with sparse rainfall during droughts or heavy rainfall during floods. And the distribution of multi-member results closely approximates a normal distribution. The normal distribution may be better able to measure the uncertainty in atmospheric processes that determine the local detail conditions. Similarly, downscaling of other atmospheric variables also show similar results (Fig. S10). Additionally, we have evaluated the spread of DPDM with 100 members (Fig. S11 and Fig. S12). Unlike many other downscaling models that tend to exhibit underestimated spread, DPDM demonstrates the capability to generate sufficient ensemble spread, even overestimating it for each variable. Furthermore, the overestimation of uncertainty suggests that DPDM can accommodate additional sources of uncertainty beyond solely the downscaling process itself.

**Application for downscaling 180-year surface climate dataset**

With the success of DPDM, we now apply the method to reconstruct high-resolution historical climate data in East Asia for the past century. Additionally, we explore several potential application scenarios for this high-resolution dataset. These application scenarios not only demonstrate the model's capability for small-scale reconstruction but also address several climate change-related issues of broad academic interest.

We select the NOAA-20C dataset as the low-resolution data because it is the only global dataset covering nearly two centuries from 1836 to 2015, encompassing the entire industrialization period. It also provides different temporal resolutions and circulation data, which not only provides the basis for constructing high-resolution surface data at six-hourly intervals but also uses the reconstructed high-resolution data with circulation to explore more on mechanisms, especially the attribution of extreme events.

To ensure the reliability of reconstruction data, we perform evaluations against the widely utilized CRU-station dataset. Given inherent data credibility issues with reanalysis datasets and the fact that observation stations solely provide precipitation and temperature data, we conduct a relative error analysis between DPDM and the Lerp results for these two datasets. It may be a crude way to evaluate our results but we do not have any other choice in the absence of high-resolution observational data for such a long period from 1836 to 2015. Fig. 4a and 4b reveal a noticeable reduction in relative errors on most stations when employing DPDM. It shows that the reconstructed data are closer to the station-observed data.

To assess the changes in aridity over the past centuries by use of the high-resolution reconstructed dataset, we have computed the Aridity Index (see Methods), a commonly applied metric with a threshold of 0.65. As shown in Fig. 4c, the shaded area represents the temporal average of AI from 2005 to 2015. The blue line, comprised of both solid and dashed segments, depicts the changes in the 11-year running mean AI at a low resolution with a constant value of 0.65. In contrast, the green line illustrates the AI values of 0.65 from 2005 to 2015 at high resolution. The findings suggest that the low-resolution data underestimates the expansion of aridity in the mid- to high-latitudes of East Asia and Northern China in the context of global warming. Fig. 4d further quantifies the area proportion of drought regions (65°E-135°E, 33°N-55°N) on a decade-by-decade basis. Remarkably, even when accounting for the uncertainty among different ensemble members, the low-resolution data persistently underestimates drought areas by approximately 3%.

Regarding the warming and humidification trend in the northwestern China during recent decades, we examine the precipitation changes in northwestern China (75°E-110°E, 30°N-50°N). The analysis reveals an actual increasing trend in precipitation since 1970, but the low-resolution data significantly overestimates precipitation intensity at a rate of 0.79 mm/day per 10 years. In contrast, the high-resolution data offers finer estimations and provides uncertainty estimates for assessing credibility, indicating a trend of 0.69 mm/day per 10 years within an uncertainty range of 0.65-0.78 mm/day per 10 years. In addition, the precipitation distribution histogram in rainy season (MJJAS) found that our high-resolution data can correctly understand the differences in climate characteristics in different regions. It tends to obtained less precipitation in drought regions while more precipitation in humid regions than low resolution (Fig. S13).

High-resolution data can provide important details for assessing extreme events. Therefore, we also conduct a simple evaluation of extreme hot and dry compound events (Figs 4e and 4f, Methods). In North China, the high-resolution reconstructed data clearly provides greater detail, detecting more extreme events and offering a finer feature of areas that are prone to such events. Additionally, it augments the available samples for subsequent attribution and synthesis analyses. Note that, while wind speed data lacks observational records, the climate statistics indicate that DPDM results also yield more detailed characters (Fig. S14). In previous studies[49], the trend in wind power change under the global warming is often estimated with the low-resolution data. However, the high-resolution data notably reveals several regions undergoing faster changes (Figs. 4g-i).

In summary, the high-resolution data generated by the DPDM not only exhibits a certain level of credibility but also enhances our understanding of climate details. This high-resolution dataset, covering the past centuries, may provide important details for improved understanding of the historical climate change.

**DISCUSSION**

In this study, we have introduced a novel probabilistic downscaling model, DPDM, for climate downscaling. We evaluate the downscaling capabilities of DPDM, which not only accurately simulates the probability distribution function of high-resolution data, but also generates a large number of members to quantify the uncertainty of the downscaled information. The latter is important since small-scale conditions under a large-scale background are never deterministic. In addition, the downscaling framework of DPDM has great potential in medium-term weather forecasts, climate predictions and future scenario projections. For instance, to generate an adequate number of ensemble members, it can be used to emulate traditional methods like the single model initial-condition large ensemble for identifying and robustly sampling extreme events [50,51].

It is undeniable that DPDM still has more room for improvement. Introducing additional circulation conditions and external forcing could enhance the model ability with more physical constraints[52]. This new approach holds promise for applications in bias correction and downsizing of dynamical model predictions. As for high-resolution climate datasets over the past centuries, while it offers valuable insights and applications, the existing datasets are limited in terms of the number of available variables and their temporal resolution. With sufficient computing resources, there is potential for increasing the temporal resolution to 6-hour intervals and downscaling other surface or upper-atmosphere variables to enable more comprehensive analyses. Additionally, we have noticed that NOAA-20CR provides corresponding ensemble spread data, which can effectively quantify the uncertainty in the initial conditions. In the future, we could sample from this ensemble spread to generate ensemble members that incorporate data uncertainty, and then apply the DPDM separately for each member. This approach would produce a larger ensemble distribution, incorporating not only the uncertainties from the downscaling process but also the uncertainties from the initial conditions. These may be reserved for future study.

In addition, the probabilistic essence and robust mathematical foundation of the diffusion model may open up a wealth of new possibilities for its practical applications in climate science as a promising tool. Its applicability extends far beyond downscaling; it holds potential for forecasting, assimilation, data reconstruction, model bias correction, sensitivity experiments, scientific inquiry, and even causal analysis. We believe that the time has come to explore the applications of the diffusion model, tackling intriguing scientific questions and contributing to the advancement of climate science.

# TABLE and FIGURES

**Table 1.** Evaluation of the downscaling performance of five surface variables from 2016 to 2021using Root Mean Square Error (RMSE), Anomaly Correlation Coefficient (ACC), Structure Similarity Index Measure (SSIM) and Peak Signal-to-Noise Ratio (PSNR) based on three different downscaling methods, including linear interpolation (Lerp), deterministic model (EDSR-GAN) and Diffusion Probabilistic Downscaling Model (DPDM) with different numbers of members. All differences are statistically significant at the 95% confidence level. Bold font highlights the best performance metrics.

| RMSE /ACC/ SSIM/ PSNR | $U_{10m}$ (m/s) | $V_{10m}$ (m/s) | $T_{2m}$ (°C) | SP (Pa) | TP (mm/day) |
|---|---|---|---|---|---|
| Lerp | 0.452/0.890/0.903/31.04 | 0.432/0.528/0.918/31.16 | 2.524/0.879/0.885/29.35 | 3191.654/0.822/0.876/24.998 | 1.197/0.8460/0.925/34.087 |
| EDSR-GAN | 0.203/0.864/0.981/**39.011** | 0.199/0.842/0.980/**39.247** | 0.867/0.880/0.987/29.698 | **106.178**/0.887/0.99968/**53.448** | 0.970/0.847/0.729//34.912 |
| DPDM single | 0.398/0.692/0.941/33.156 | 0.457/0.601/0.905/28.083 | 1.295/0.785/0.936/32.585 | 159.406/0.726/0.99943/51.070 | 2.670/0.679/0.569/26.242 |
| DPDM 33 | 0.185/0.901/0.983/34.37 | 0.193/0.877/0.981/34.234 | 0.677/0.943/0.992/38.268 | 108.225/0.892/0.99969/52.073 | 0.915/0.872/0.948/37.075 |
| DPDM 100 | **0.175**//**0.911**/**0.984**/33.322 | **0.181**/**0.895**/**0.984**/37.085 | **0.657**/**0.947**/**0.994**/**39.256** | 106.706/**0.899**/**0.99969**/51.065 | **0.886**/**0.878**/**0.948**/**37.095** |

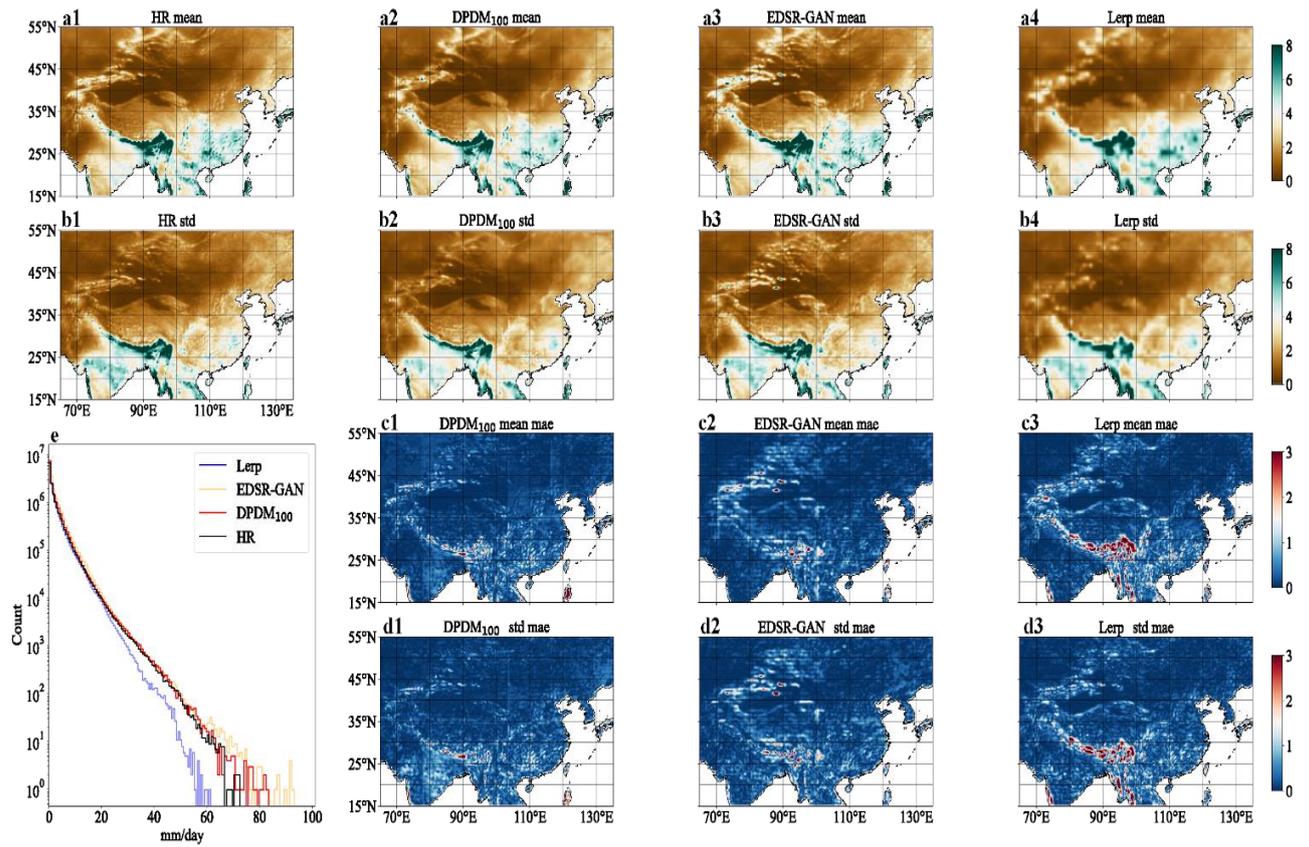

**Figure 1. The downscaling results for climatological mean total precipitation and its standard deviation based on different methods.** (a) Observed and downscaled climatological mean precipitation during 2016-2021 based on the three different methods. (b) As in (a), but for the standard deviation of monthly total precipitation. (c) Bias of the downscaling results relative to the true high-resolution precipitation based on the different methods. (d) As in (c), but for the bias in variability (standard deviation), between the downscaling results and the high-resolution data. (e) Total precipitation distribution histograms based on the high-resolution (black curve), EDSR-GAN (yellow curve), DPDM with 100 members (red curve) and Lerp (blue curve).

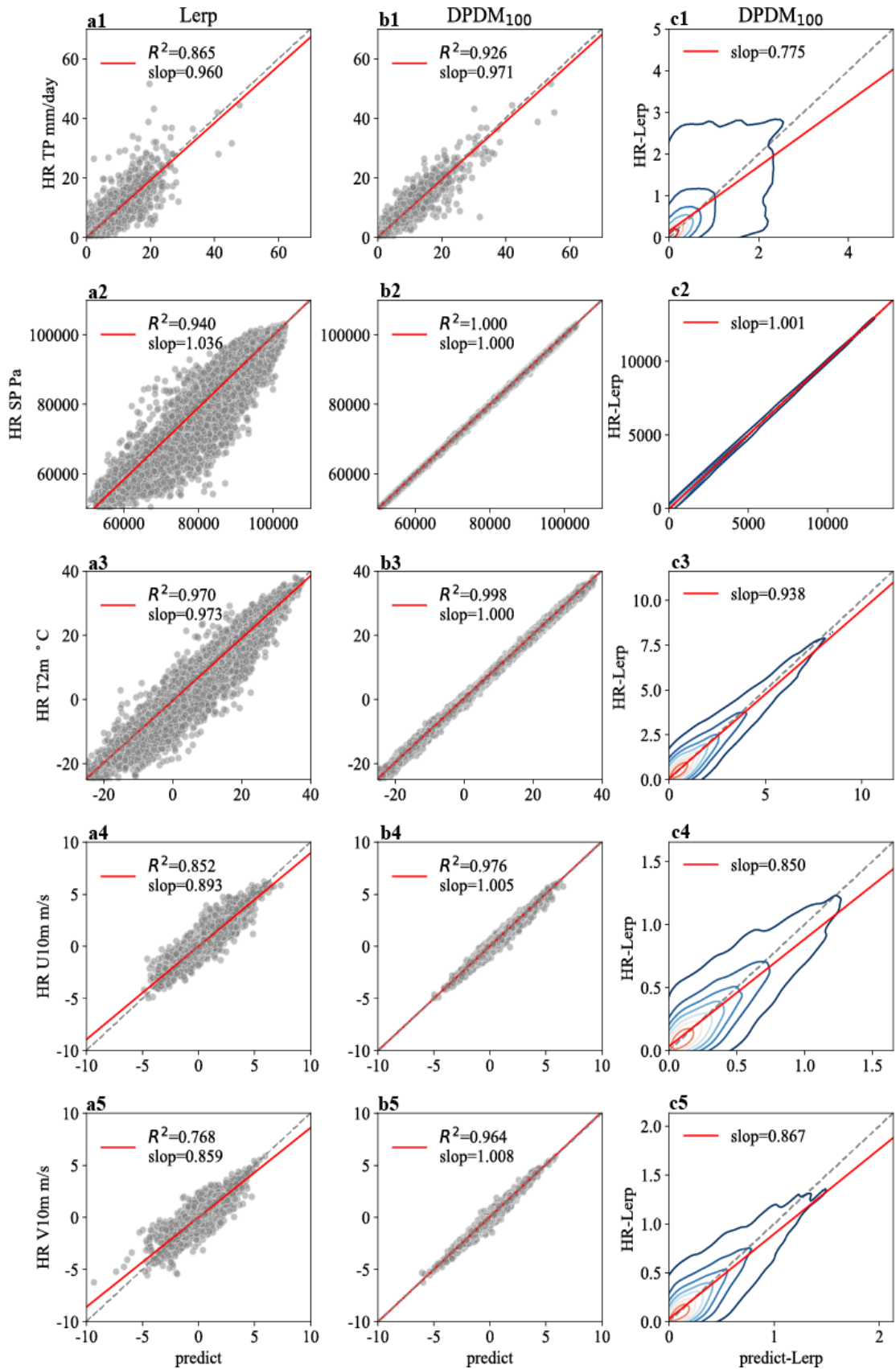

**Figure 2. Evaluation of R squared ($R^2$) and blurriness for downscaling results in**

**different variable.** (a1-a5) Scatter plots between the high-resolution observations and the downscaling results with the Lerp. The red line represents the fitting line, and slope and $R^2$ are shown in the legend. (b1-b5) As in (a1-a5), but for the results of DPDM with 100 members. (c1-c5) Two-dimensional kernel density estimation of DPDM with 100 members to assess the blurriness of downscaling results. The x axis represents the absolute difference between the downscaling outputs and the Lerp, and the y axis represents the absolute difference between the high-resolution observations and the Lerp. The black dashed diagonal line represents the identity relation.

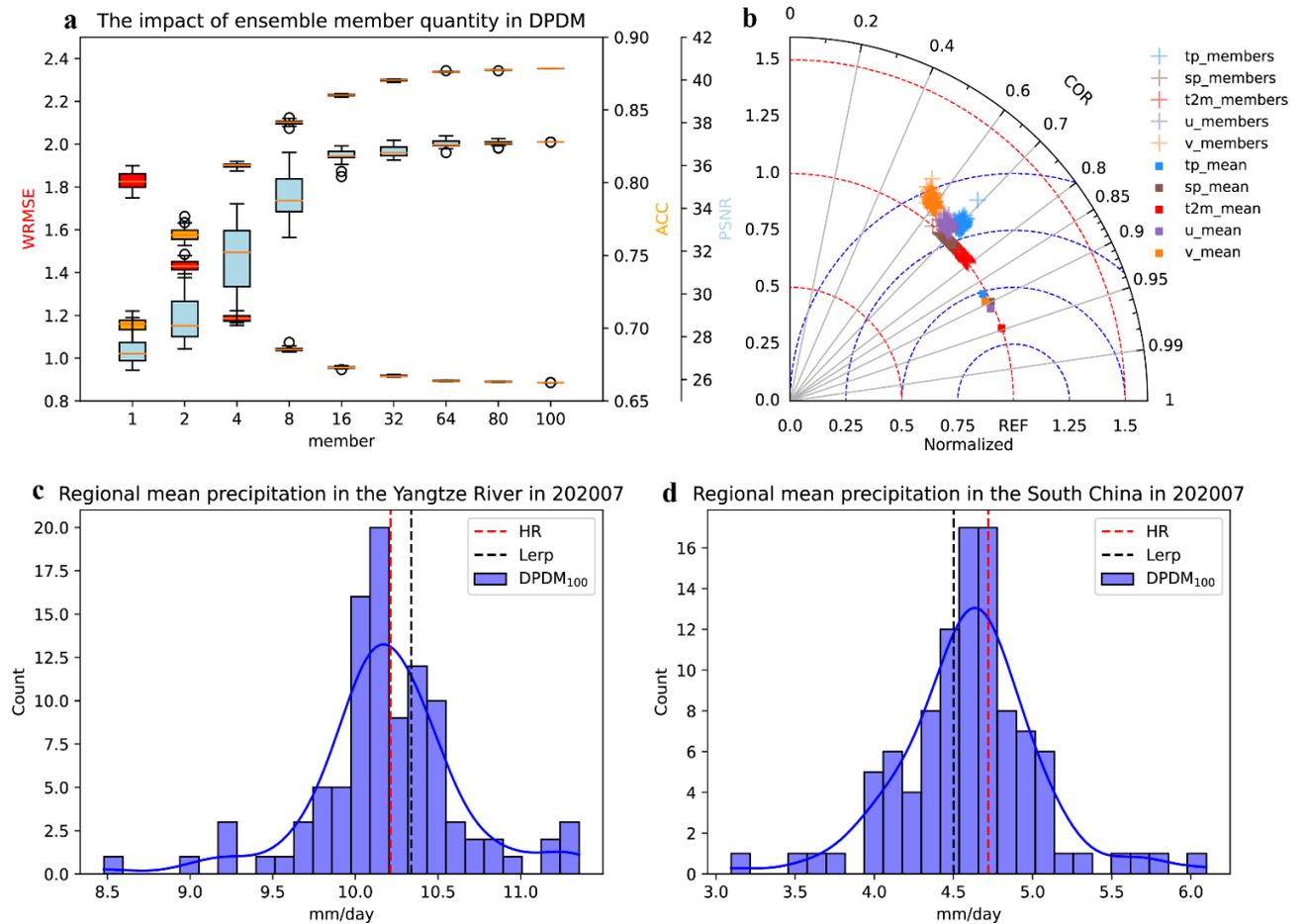

**Figure 3. The role of different members in the DPDM.** (a) Impact of ensemble member quantity on RMSE, ACC and PSNR of precipitation downscaling performance with increasing number of ensemble members. Each subset based on non-repetitive random sampling of 'n' members from the existing pool of 100 members. To mitigate random errors, this resampling process is repeated 20 times. (b) Taylor plots of different variables for comparisons between individual members and the multi-member ensemble mean. Different colors represent different meteorological variables, crosses represent individual members, and squares represent multi-member ensemble averages. (c) Regional mean precipitation from different members of DPDM in the Yangtze River Basin (24°-35°N, 90°-123°E) in July 2020. The red line represents the real high-resolution result, the black dashed line indicates the Lerp, the blue line denotes the probability density distribution of the 100 members and the blue line denotes its fitting curve. (d) As in (c), but for the results of regional mean precipitation in South China (18°-26.5°N, 90°-123°E) in July 2020.

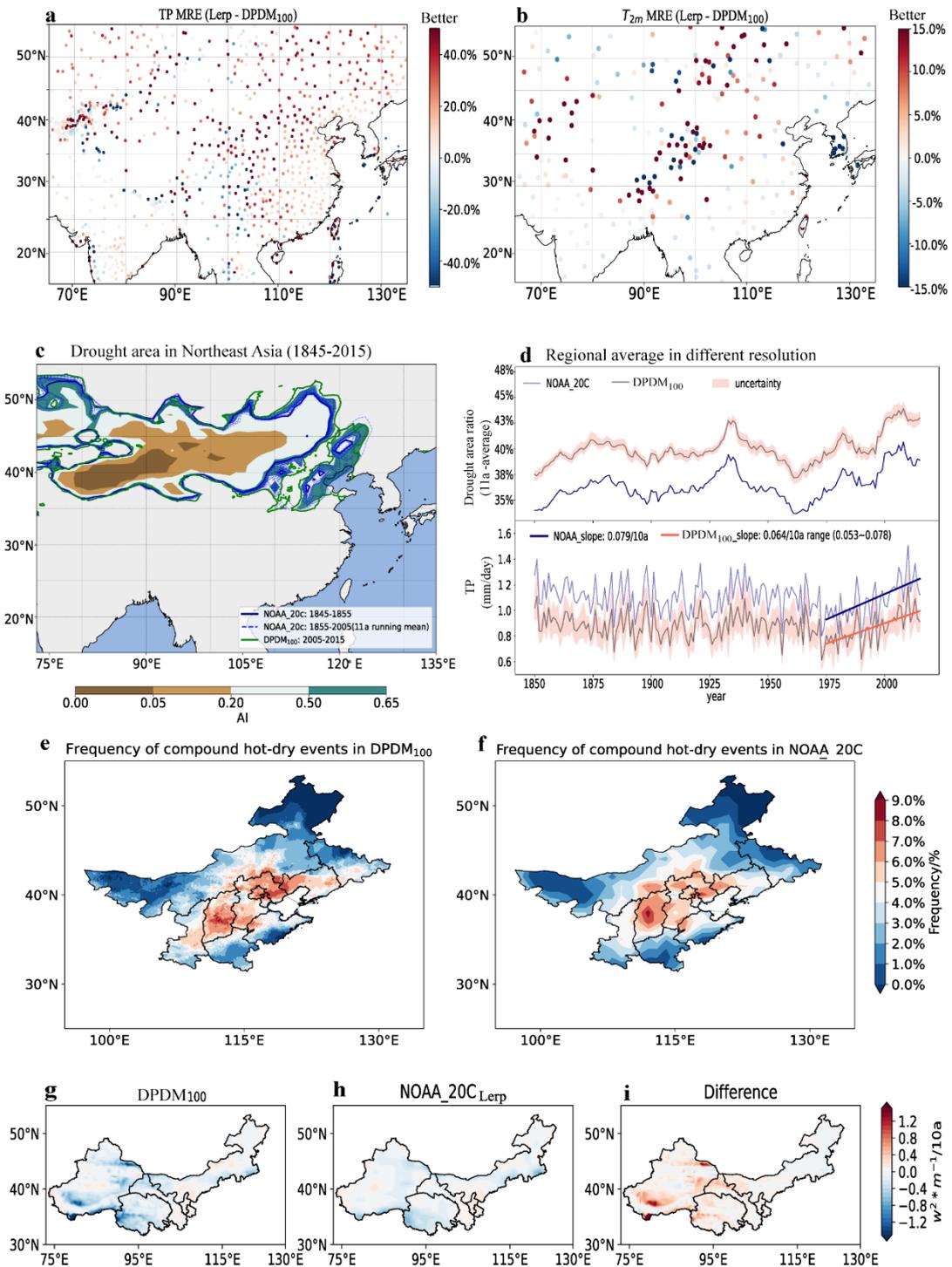

**Figure 4. Applicable scenarios for high-resolution datasets over the past 180 years using DPDM.** (a) Comparison of relative errors of the Lerp and DPDM relative to station observations. The red points indicate that the DPDM has a lower error than the Lerp in precipitation downscaling, and the blue points indicate the opposite. (b) As in

(a), but for temperature at 2 meters high. (c) Distribution of drought area in mid-high latitudes of East Asia. The shading represents the temporal average of Aridity Index (AI) from 2005 to 2015 at low resolution. The blue solid line represents the AI of 0.65 from 1845 to 1855 and the blue dashed line represents the 11-year running mean AI of 0.65 from 1855 to 2005 at low resolution. The green line represents the AI of 0.65 from 2005 to 2015 at high resolution. (d) Comparison of regional average based on high- and low-resolution datasets. The upper panel shows the proportion of dry areas in the middle and high latitudes of East Asia (33°-55°N, 65°-135°E) at different resolutions. The lower panel shows seasonal mean precipitation in the rainy season (May-September) and the precipitation change trend during the past 40 years in the northwest China (35°-50°N,70°-110°E) at different resolutions. The red line indicates the ensemble mean result of DPDM, and red shading gives the uncertainty among the DPDM members. The blue line represents the result of the low-resolution NOAA-20C. (e) Frequency of composite hot-dry events in North China over the past 180 years from the high-resolution reconstructed data. (f) As in(e), but for the results of NOAA-20C. (g) Change trend of wind power in Xinjiang from 1960 to 2015 based on the high-resolution reconstructed data. (h) As in (g), but for the results of NOAA-20C. (i) Differences in wind power change trends in Xinjiang between the different resolutions from 1960 to 2015.

# METHODS
## Diffusion Probabilistic Downscaling model

Denoising diffusion probabilistic model has become escalatingly influential in recent years. It is a type of generative models inspired by considerations from nonequilibrium thermodynamics[53], which defines a bidirectional Markov chain with length $T$. The forward diffusion process gradually adds Gaussian noise to the original input data $x_0$, generating a sequence of data $x_1 \dots x_T$. The reverse diffusion process iteratively removes noise from the noisy image $x_t$ by sampling from $p(x_{t-1}|x_t)$. To generate new data sample, it samples Gaussaian noise map $x_T$ from the normal Gaussian distribution $N(0, \sigma_{max}^2 I)$ as the input of the model. Then, the model $\epsilon_\theta$ predicts the noise $\epsilon_\theta(x_t, t)$ based on the current state $x_t$ in each diffusion step t to gradually removes noise, thus reversing the diffusion process.

DPDM is a variant of the conditional diffusion probabilistic model with low-resolution input, topography, and land-sea mask as conditions called as $\tilde{x}$. DPDM removes noise from the noisy image $x_t$ by sampling from the joint probability $p(x_{t-1}|x_t, \tilde{x})$. The DPDM model is trained with the L2 loss between the predicted noise and the actual noise $||\epsilon - \epsilon_\theta(x_t, t, \tilde{x})||_2^2$. Here, we give a brief introduction to DPDM in forward and reverse process (more details in Ho et al [33,38,39,54]).

## Forward Diffusion process

The forward process of DPDM is similar to iteratively constructing a mapping relationship from a high-resolution distribution to pure gaussian distribution using the Markov chain with total length $T$. More specifically, given a data sampled from the real high resolution data distribution $x_0 \sim q(x)$, the forward diffusion process, in which we gradually add Gaussian noise to $x_0$ in total steps ($T$) with the noisy density for each step ($t$) being controlled by a variance schedule $\beta_1 \dots \beta_T$, produces the sequence of high-resolution data with noisy samples $x_1 \dots x_T$.

$$q(x_t|x_{t-1}) = N(x_t; \sqrt{1-\beta_t}x_{t-1}, \beta_t I) \tag{1}$$

If the magnitude $\beta_t$ of the noise added at each step is small enough, and the total step $T$ is large enough (in our experiments, $T$ is set to 1000 steps), then $x_T$ is equivalent to an isotropic Gaussian distribution $x_T \sim$ N (0, I). With the help of the properties of Markov chain, we successfully connect the high-resolution data distribution $q(x)$ with the Gaussian distribution N (0, I). Then, we can obtain the noise samples at step t in the forward process, using the following equation:

$$q(x_{1:T}|x_0) = \prod_{t=1}^{T} q(x_t|x_{t-1}) = \prod_{t=1}^{T} N(x_t; \sqrt{1-\beta_t}x_{t-1}, \beta_t I) \tag{2}$$

In order to obtain $x_t$ without iteration for fast training, we can further expand $x_t$ with the help of reparameterization trick and additive nature of Gaussian distributions. Let $\alpha_t = 1 - \beta_t$ and, $\bar{\alpha}_t = \prod_{i=1}^{t} \alpha_i$, we can get the $x_t$ and $q(x_T|x_0)$:

$$x_t = \sqrt{\alpha_t}x_{t-1} + \sqrt{1-\alpha_t}\epsilon_{t-1} = \sqrt{\bar{\alpha}_t}x_0 + \sqrt{1-\bar{\alpha}_t}\epsilon, \quad \epsilon \sim N(0, I) \tag{3}$$

$$q(x_T|x_0) = N\left(\sqrt{\bar{\alpha}_t}x_0, (1-\bar{\alpha}_t)I\right) \tag{4}$$

**Reverse Diffusion process**

Now we can conveniently sample $x_t$ at an arbitrary timestep t in the forward process. At the timestep T, we can sample gaussion noise map $x_T$ from the gaussion distribution $N(0, I)$. Then we need to estimate $q(x_{t-1}|x_t)$ to remove the gaussion noise added in the forward process of DPDM. Unfortunately, we cannot easily estimate $q(x_{t-1}|x_t)$ because it requires to use the entire dataset. Therefore, we learn the DPDM model $p_\theta$ conditioned on the state $x_t$ at the timestep $t$ and the conditional data $\tilde{x}$ to approximate these conditional probabilities for getting the next state $x_{t-1}$.

$$p_\theta(x_{t-1}|x_t, \tilde{x}) = q(x_{t-1}|x_t, x_0) = N\left(x_{t-1}; \mu_\theta(x_t, t, \tilde{x}), \Sigma_\theta(x_t, t)\right) \quad (5)$$

Through Bayes' rule, we can get the following equation (6) and (7). Subsequently, based on the Markov assumption, equation (8) can be obtained from (7). Finally, by utilizing Gaussian distribution functions from the forward process, all components of equation (8) can be expressed, simplifying it to equation (9):

$$q(x_{t-1}|x_t, x_0) = \frac{q(x_{t-1}, x_t, x_0)}{q(x_t, x_0)} \quad (6)$$

$$= \frac{q(x_t|x_{t-1}, x_0)q(x_{t-1}|x_0)q(x_0)}{q(x_t|x_0)q(x_0)} = \frac{q(x_t|x_{t-1}, x_0)q(x_{t-1}|x_0)}{q(x_t|x_0)} \quad (7)$$

$$= \frac{q(x_t|x_{t-1})q(x_{t-1}|x_0)}{q(x_t|x_0)} \quad (8)$$

$$\approx \exp\left(-\frac{1}{2}\left(\left(\frac{\alpha_t}{\beta_t} + \frac{1}{1-\bar{\alpha}_{t-1}}\right)x_{t-1}^2 - \left(\frac{2\sqrt{\alpha_t}}{\beta_t}x_t + \frac{2\sqrt{\bar{\alpha}_{t-1}}}{1-\bar{\alpha}_{t-1}}x_0\right)x_{t-1} + C\right)\right) \quad (9)$$

Following the standard Gaussian distribution function, the mean and variance can be parameterized as follows:

$$\mu_\theta(x_t, t) = \frac{1}{\sqrt{\alpha_t}}\left(x_t - \frac{1-\alpha_t}{\sqrt{1-\bar{\alpha}_t}}\epsilon_\theta(x_t, t, \tilde{x})\right), \Sigma_\theta(x_t, t) = \frac{\beta_t(1-\bar{\alpha}_{t-1})}{1-\bar{\alpha}_t}I \sim \beta_t I \quad (10)$$

According to equation (9) and (10), we can obtain the distribution function (11), and we can sample it to get the $x_{t-1}$ thereby realizing the reverse denoising process.

$$x_{t-1} = N\left(x_{t-1}; \frac{1}{\sqrt{\alpha_t}}\left(x_t - \frac{1-\alpha_t}{\sqrt{1-\overline{\alpha_t}}}\epsilon_\theta(x_t, t, \tilde{x})\right), \sqrt{\beta_t}\epsilon\right)$$

$$= \frac{1}{\sqrt{\alpha_t}}\left(x_t - \frac{1-\alpha_t}{\sqrt{1-\overline{\alpha_t}}}\epsilon_\theta(x_t, t, \tilde{x})\right) + \sqrt{\beta_t}\epsilon \quad \epsilon \sim N(0, I) \quad (11)$$

**Training details and model structure**

To obtain a sufficient volume of training data, we employ a dataset with a 6-hour time resolution instead of the monthly dataset. We established a connection between the ERA5 product (1° spatial resolution), and the high-resolution ERA-land dataset (0.1° spatial resolution). These reanalyses are provided by the Copernicus Climate Change Service at ECMWF, combining a large range of satellite-based and land-based observations with high-resolution model simulations through state-of-the-art data assimilation techniques, spanning the period from 1961 to 2021.

In this study, we split the ERA-Land and ERA5 datasets into the training period of 1960–2015 and the test period of 2015–2021. In order to consider the effect of terrain, we use the topography, land-sea mask and low-resolution data information as the input of DPDM for training. We randomly crop 128*128 patch from the original low-resolution data and get the corresponding high-resolution data for training, instead of using the original low- and high-resolution data for training. The reason why we use this strategy is because, this can not only increase the diversity of data to learn the mapping of low-resolution data to high-resolution data, but also reduce the consumption of computing resources, such as graphics memory and computational cost. With the ablation study (Fig. S15), we investigated the effects of adding terrain information to the patch input and explored the impact of different patch sizes (32,64,128,256) on model performance. Our experiments revealed that incorporating terrain information led to faster loss convergence, patch size of 128 can obtained optimal performance and patch size of 256 resulted in non-converging loss.

Through the patch strategy, the model can dynamically process low-resolution data to generate high-resolution data for any region and condition, instead of just fitting in a certain region, thereby improving the generalization ability of the model. For example, it can obtain high-resolution results outside the target region without the training, albeit with limited effectiveness, as seen in the Gulf of Mexico (Fig. S16) and northern Australia (Fig. S17). It should be noted that in the DPDM, we also need to use the above patch strategy for inference, rather than using the entire low-resolution data. The input image is divided into different patches, each patch is processed individually, and then the results are combined to obtain the final output. To ensure continuity at the boundaries, we introduce overlapping between some patches. However, there still be some quality degradation at the boundaries of patches. Increasing the amount of overlap or employing advanced techniques for boundary restoration can yield more consistent results, but this approach requires sufficient computational resources. For different surface variable, we adopt different normalization strategies to facilitate rapid loss convergence and training stability. Precipitation data undergoes a transformation by adding one and then taking the logarithm. Other variables are standardized and normalized to a range between 0 and 1. We also standardize the topography data and change the land-sea mask information to a matrix containing only 0 and 1.

The DPDM model architecture is similar to SR3, which is a U-Net-like architecture (Fig.S1). At each time step $t$, the model's input comprises the concatenation of two datasets: the conditional data $\tilde{x}$ and the noisy data $x_t$, both have the same dimensions as the high-resolution data $x_0$. Conditional data $\tilde{x}$ includes interpolation result obtained by Lerp interpolating low-resolution data, topography, and land-sea mask information. Concatenation is a simple and effective method to add the conditional data to the model. Then, a convolutional layer with 64 kennels is used to extract the input data information. Downsampling modules are applied in DPDM consisting of several residual blocks and self-attention layers[55]. Based on empirical experience, we only use 3 downsampling modules, and the data dimensions are reduced from 128 to 16. The upsampling modules is similar to the downsampling modules. All the convolutions in our model use the 3*3 convolution kernel size and the 1*1 stride. For encoding the timestep t, we use the sinusoidal positional encoder[55] that contains two fully-connected layers and a sigmoid linear unit (SiLU) activation function between the two layers. Then we add the timestep feature encoded by above sinusoidal positional encoder to our intermediate feature maps after the group normalization operators in each residual block.

**Climate index and Compound events**

An aridity index (AI[56]) is a numerical indicator of the degree of dryness of the climate at a given location.

$$AI = \frac{P}{PET} \qquad (12)$$

where PET is the potential evapotranspiration, which is calculate by the python package of Climate Indices (https://climate-indices.readthedocs.io/en/latest/) and the P is the annual average precipitation.

Compound hot-dry events are defined as the co-occurrence of high mean temperature anomaly (above the 90th percentile) and low mean precipitation anomaly (below the 10th percentile) values over the warm season[51] (i.e., the three consecutive months with highest mean temperature during 1836–2015) for each grid point, The centennial trends of the two variables are removed.

Wind energy[57] is a typical measure of wind energy potential, defined as follows:

$$Wind\ power = \frac{1}{2}\rho W_h^3 \quad (13)$$

where ρ represents the air density, which is assumed to be a constant value of $1.213\ kg * m^{-3}$ at standard atmospheric conditions, and $W_h$ is approximately expressed by the wind speed at a height of 10 m.

**Metric**

Based on the downscaling results, we compute some metrics, i.e., Anomaly Correlation Coefficient (ACC), Peak Signal-to-Noise Ratio (PSNR), Structure Similarity Index Measure (SSIM), Root Mean Square Error (RMSE), and Normalized Root Mean Square Error (NRMSE) defined as follows:

$$ACC = \sum_{m=1}^{12} \frac{\sum_{y=s}^{e}(O_{y,m} - \overline{O_m})(D_{y,m} - \overline{D_m})}{\sqrt{\sum_{y=s}^{e}(O_{y,m} - \overline{O_m})^2 \sum_{y=s}^{e}(D_{y,m} - \overline{D_m})^2}} \quad (14)$$

$$PSNR = \frac{1}{n}\sum_{i=1}^{n} 10 * log_{10}(\frac{MaxValue^2}{MSE}) \quad (15)$$

$$MSE = \frac{\sum_{j=1}^{N_{lat}} \sum_{k=1}^{N_{lon}} (D_{j,k} - O_{j,k})^2}{N_{lat} \times N_{lon}} \quad (16)$$

$$SSIM = \frac{1}{n}\sum_{i=1}^{n} \frac{(2\mu_{D,i}\mu_{O,i} + C_1)(2\sigma_{D,i}\sigma_{O,i} + C_2)}{(\mu_{D,i}^2 + \mu_{O,i}^2 + C_1)(\sigma_{D,i}^2 + \sigma_{O,i}^2 + C_2)} \quad (17)$$

$$RMSE = \sqrt{\frac{1}{n}\sum_{i=1}^{n} \frac{\sum_{j=1}^{N_{lat}} \sum_{k=1}^{N_{lon}} (D_{i,j,k} - O_{i,j,k})^2}{N_{lat} \times N_{lon}}} \quad (18)$$

$$NRMSE = \frac{\sqrt{\frac{1}{n}\sum_{i=1}^{n}(O_i - D_i)^2}}{\sqrt{\frac{1}{n}\sum_{i=1}^{n}(O_i - \bar{O})^2}} \quad (19)$$

Here, $O$ and $D$ denote the observed and the downscaling results, respectively. $\overline{O_m}$ and $\overline{D_m}$ denote the climatology of observed and the downscaling results in each calendar month $m$ (from 1 to 12). The label $y$ denotes the forecast target year. Finally, $s$ and $e$ denote the earliest (that is, 2016) and the latest year (that is, 2021) of the validation, respectively. $MaxValue$ denote the maximum value of the normalized data (that is, 1). $\mu_{D,i}$ and $\sigma_{D,i}$ represents the spatial means and standard deviations of the downscaling results, while $\mu_{O,i}$ and $\sigma_{O,i}$ represents the spatial means and standard deviations of observation. $C_1$ and $C_2$ are constants to avoid computation instability when the denominator approaches zero.

**ACKNOWLEDGEMENTS**

This work is supported by the National Key Research and Development Program of China (No. 2020YFA0608000), National Natural Science Foundation of China (Grant 42030605) and the program of China Scholarships Council (No. CXXM2101180001). We gratefully acknowledge the support from the Huawei MindSpore team. The code and dataset will be released on the MindSpore platform.

**AUTHOR CONTRIBUTIONS**

F.H.L. and Z.L. are co-first authors and wrote the manuscript. F.H.L. and Z.L. designed the AI models. F.H.L. prepared for data and performed the main experiments. Z.L. reconstructed long-term East Asian climate historical datasets. F.H.L., Z.L., L.B. performed the analysis under supervision of J.-J.L. J.-J.L., S.B., D.J., B.P. and T.Y. conducted analysis from the climate science view. All authors contributed to interpreting results, discussions of associated dynamics and improvement of the presentation.

**COMPETING INTERESTS**

All authors declare no competing interests.